\documentclass{iaus}
\usepackage{graphicx}

\title[ Extragalactic GCs in the near-infrared] 
{An optical/near-infrared survey of GCs in early-type galaxies}

\author[Ana L. Chies-Santos et al.]   
{A. L. Chies-Santos$^1$,
S. S. Larsen$^1$, E. M. Wehner$^1$, H. Kuntschner$^2$, \\J. Strader$^3$,  J. P. Brodie$^4$ and J. F. C. Santos Jr$^5$}

\affiliation{$^1$Sterrenkundig Instituut, University of Utrecht, The Netherlands, 
 $^2$ST-ECF/ESO, Germany $^3$Harvard CfA, USA, $^4$UCO/Lick Observatory, USA, $^5$UFMG, Brazil
\\email: a.l.chies@uu.nl}

\pubyear{2010}
\volume{S266}  
\pagerange{1--6}
\jname{Star Clusters: Basic Galactic Building Blocks Throughout Time And Space}
\editors{Richard de Grijs \& Jacques Lepine, eds.}
\begin{document}

\maketitle

\begin{abstract}
Optical/near-infrared observations for 14 globular cluster (GC) systems 
in early-type galaxies are presented. 
We investigate the recent claims (\cite[Yoon, Yi \& Lee 2006]{yoon06}) of colour bimodality in GC systems
being an artefact of the non linear colour - metallicity transformation driven by the
horizontal branch morphology. 
Taking the advantage of the fact that the combination of optical and near-infrared 
colours can in principle break the age/metallicity degeneracy we also analyse age distributions in these systems.

\end{abstract}

\firstsection 
\section{Introduction}

The ubiquity of globular clusters (GCs) around all major galaxies and the fact that young star clusters are being observed in many mergers and starbursts we observe today suggests that GC formation traces important star formation events in the history of their host galaxies (\cite[Brodie \& Strader 2006]{bs06}). 

GC systems generally exhibit a colour distribution that is bimodal (sometimes multimodal) in optical colour. 
Spectroscopic studies (eg. \cite[Strader et al. 2005]{strader05}, \cite[Cenarro et al. 2007]{cenarro07}) show that this bimodality is due to old subpopulations with ages greater than 10 Gyrs that differ in metallicity. 
Bimodal metallicity distributions may indicate distinct formation mechanisms, with age differences $\sim2$\,Gyrs still being allowed within the uncertainties on current age estimates.
This metallicity bi/multimodality seems to hold for the majority of galaxies, except for the low mass galaxies which appear to have only a metal poor GC subpopulation.
Recently \cite[Yoon, Yi \& Lee (2006)]{yoon06} challenged this metallicity bimodality interpretation saying that it is an artefact of the horizontal branch morphologies. These authors argue that non-linear colour metallicity relations may transform a unimodal metallicity distribution into a bimodal optical colour distribution. This issue has been investigated in more detail by \cite [Cantiello \& Blakeslee (2007)]{cb07} who concluded that combinations of optical and NIR colours are much less prone to this effect. 
Metallicity bimodality was thus contested and it is therefore important to investigate this topic in more detail. On the other hand, if metallicity bi/multimodality is genuine there could be multiple episodes/mechanisms of GC formation in a galaxy with a possible age difference. It is thus equally important to study the age distributions of these systems.
It is still an open issue in the literature whether some early type galaxies known to contain old stellar populations from integrated light studies can host a high percentage of intermediate age (2-3 Gyrs) GCs. The two strongest cases are NGC\,4365 (\cite[Puzia et al. 2002]{puzia02}, \cite[Larsen et al. 2005]{lbs05}) and NGC\,5846 (\cite[Hempel et al. 2003]{hempel03}). The comparison of 2-colour plots of the GC systems with simple stellar population (SSP) models (eg. \cite[Bruzual \& Charlot 2003]{bc03}, \cite[Maraston 2005]{m05}) suggested the presence of intermediate age GCs in these galaxies.  

With the aim of investigating ages and metallicity distributions in GC systems of early-type galaxies we examine 14 E/S0 systems, with $(m-M)<32$ and $M_B$  $<$ $-19$ through optical and NIR imaging. We obtained deep ($\sim 3.4$ hours) K-band photometry with the LIRIS instrument in the William Herschel Telescope in La Palma which has a field of view of $4.2\,\times\,4.2\,arcmin^{2}$. These were combined with optical imaging extracted from the Hubble Space Telescope public archive in $g$ (F475W) and $z$ (F850LP) bands.  
GC candidates were selected automatically in galaxy subtracted images through standard routines. The following cuts were applied, defining the final sample: $g$ magnitude  $g\,<\,23$, optical colour $0.5\,<\,(g-z)\,<\,2$, effective radii $1\,<\,R_{eff}\,<\,15\,pc$ and magnitude error $K_{err}\,<\,0.5$. With these restrictions, typical magnitude errors for the optical bands $g$ and $z$ came to be $\sim\,0.05$. For the NIR band the magnitude errors were measured to be $\sim\,0.06$ for a $K=19$ GC and $\sim\,0.2$ for a $K=20$ GC. Artificial star tests though show that the $K$-band magnitude errors can be underestimated by a factor of 2.
In what follows it is shown what knowledge was gained from this data regarding the age distributions and the colour (metallicity) distributions of the GC systems studied here.

\section{Ages}
The combination of optical/NIR imaging is an alternative to spectroscopy in age dating GCs. The technique relies on comparing 2-colour diagrams with the predictions of SSP models.
In Fig.\,1 2-colour diagrams $(g-k)$ \textit{vs.} $(g-z)$ for the GCs that satisfy the criteria mentioned above for the 14 galaxies are shown.
In the left panel, a model grid from \cite[Maraston (2005)]{maraston05} is overplotted with model sequences of constant age as solid lines: 2, 3, 4, 5, 6, 8, 11, 14 Gyrs.
If one examines the case of NGC\,4365, the galaxy that showed evidence for having intermediate age clusters one concludes that this galaxy is no different from any other in the sample.
In the right panel the same data as the left panel is shown but now over ploted with Padova SSPs with \cite[Marigo et al. (2008)]{marigo08} isochrones with a new treatment of the  thermally pulsing asymptotic giant branch (TP-AGB) phase with the same age range as for the first panel. These latter models for old ages ($\sim 11, 14$ Gyrs) are a close match to the data, while \cite[Maraston (2005)]{maraston05} models show an offset that would imply intermediate ages ($\sim2, 3$ Gyrs). \cite[Charlot \& Bruzual (2009)]{cb09} preliminary models yield basically the same conclusion as the \cite[Maraston (2005)]{maraston05} models: the $(g-k)$ SSPs are $\sim\, 0.2-0.3$ mag too blue. A similar offset is also the case for all older SSP models.
The only indicative for intermediate ages is the presence of half a dozen GCs with $\sim(g-k)\,\sim\,3.7$ in NGC\,4406.  
The data for NGC\,4382 and NGC\,4473 GC systems should be taken with caution as they were observed in non-photometric conditions.
The data spread along $(g-k)$ is attributed to observational uncertainties or is something intrinsic of the systems. For the case of M87, the GCs present a standard deviation of $\pm0.75$ along $(g-k)$ from the best fit line between $(g-k)$ and $(g-z)$ and the observational errors reach a maximum of $\pm0.4$ magnitudes (according to artificial star tests).  
One might think, [$\alpha$/Fe] ratios could maybe account for part of this spread, but this of course would only explain the scatter if the ratios varied among clusters.
Stochastic effects in the integrated colours are investigated following \cite[Santos \& Frogel (1997)]{sf97}. It is found that the fluctuations in the number of bright stars can account for only $\sim\,0.1$ magnitudes of the spread in $(g-k)$ for a 14 Gyr old SSP and a $\sim\,0.2$ for a 2 Gyr one assuming that all clusters have $10^{6}$ stars. 
We conclude that age dating GC systems as old ($\sim 11, 14$ Gyrs) or intermediate age ($\sim 2, 3$ Gyrs) depends on the choice of the SSP model.
Padova SSPs with \cite[Marigo et al. (2008)]{marigo08} isochrones for old ages are in good agreement with the data. Moreover there is no evidence for significant differences in the GC age distributions among the galaxies studied here. NGC\,4406 could be an exception.
 
\begin{figure}[h]
\begin{center}
\includegraphics[width=1.9in]{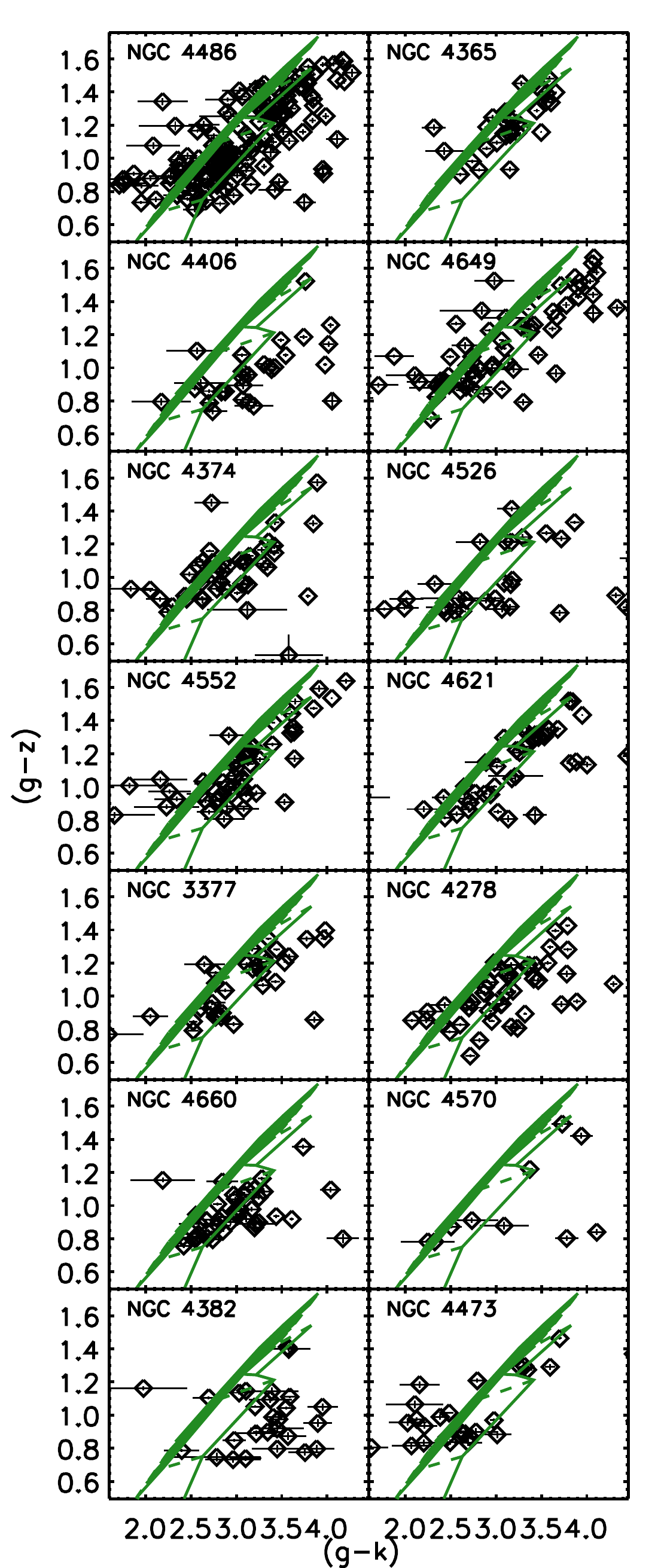} 
\includegraphics[width=1.9in]{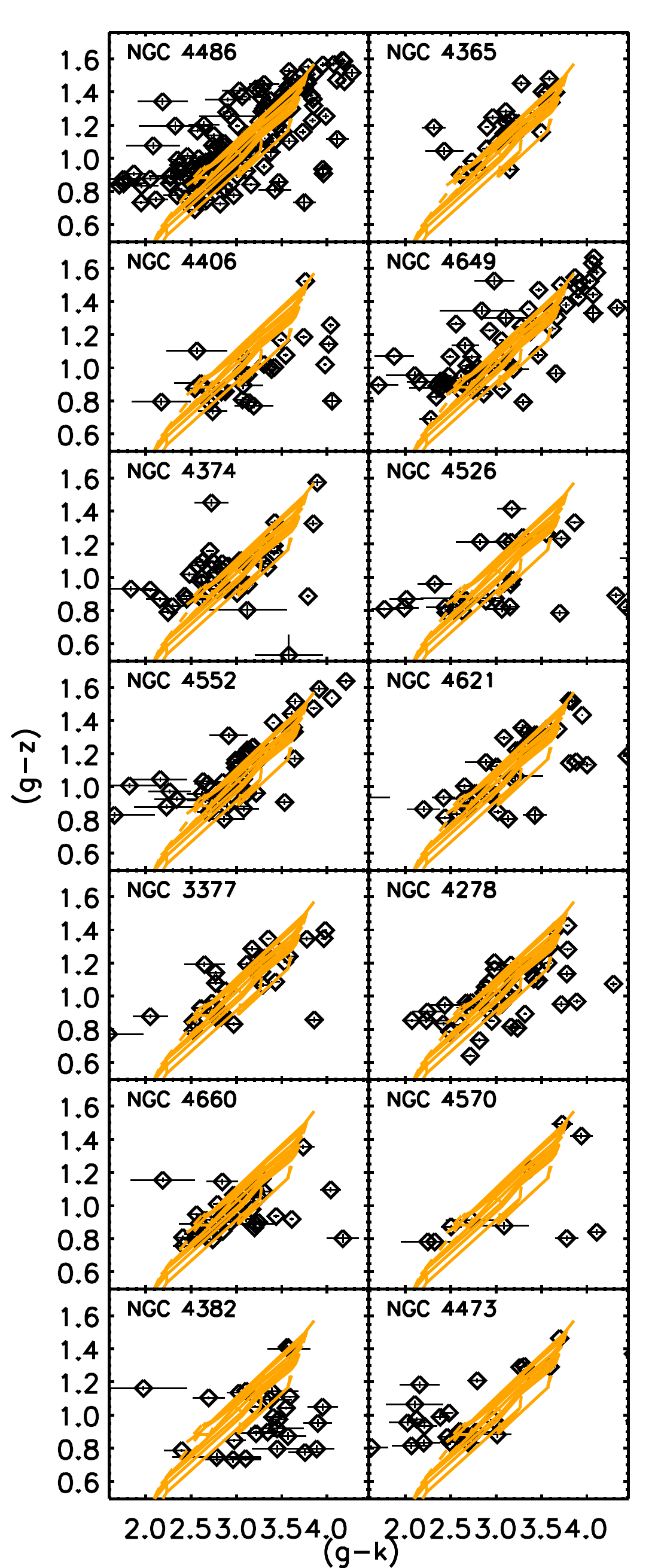} 
 \caption{2 colour diagrams for the GC systems of the 14 galaxies. $(g-k)$ \textit{vs.} $(g-z)$.
 \textit{Left panel}: A model grid from \cite[Maraston (2005)]{m05} is overplotted with model sequences of constant age as solid lines: 2, 3, 4, 5, 6, 8, 11, 14 Gyrs. 
  \textit{Right panel}: A model grid of Padova SSPs with \cite[Marigo et al. (2008)]{marigo08} isochrones is now overplotted with the same ages as in the left panel.}
   \label{Fig1}
\end{center}
\end{figure}

\section{Bimodality}
The combination of optical and NIR colours is a better indicator of the true underlying  
metallicity distribution than optical colours alone (\cite[Cantiello  
\& Blakeslee 2007]{cb07}). At redder  
wavelengths the stellar population will have decreasing contributions  
from horizontal branch stars.
If the metallicity distribution of a GC system is genuinely bimodal, this as seen through optical colours should be maintained as one moves to redder wavelengths.  
We thus test the alternative \cite[Yoon, Yi \& Lee (2006)]{yoon06} scenario for colour bimodality.

In Fig.\,2. the $(g-z)$, $(g-k)$ and $(z-k)$ colour distributions for the GC systems of the 14 E/S0s shown in Fig.\,1. are plotted in the first, second and third panels respectively. Note that the bimodality becomes less evident in $(g-k)$ if compared to $(g-z)$ and even less pronounced in $(z-k)$. This is the colour that should have the smallest contribution from horizontal branch stars. It is evident from this figure that the galaxies that have more than one peak in $(g-z)$ appear more strongly single peaked in $(z-k)$ while the multiple peaks in the $(g-k)$ distributions still appear more perceptive. 
Note how this effect is pronounced in NGC\,4486 (M\,87) the most cluster rich galaxy in the sample. 
This result fits very nicely into the \cite[Yoon, Yi \& Lee (2006)]{yoon06} picture where the horizontal branch morphology would induce a nonlinear behaviour in the colour-metallicity relation and this would drive a unimodal metallicity distribution into a bimodal (optical) colour distribution. If in the reddest of the available colours $(z-k)$, the best metallicity indicator available here the distribution is unimodal it seems straightforward to conclude that the horizontal branch morphology could be causing this.
Taking into account that $(z-k)$ is the most sensitive colour, among the ones available in this study, to uncertainties in the K-band we tested whether this observational scatter could blur the bimodality. We carried out a linear transformation of the $(g-z)$ colour distribution to the $(g-k)$ and $(z-k)$ ones that preserved bimodality in the absence of noise. 
We then simulated blue and red peaked gaussian distributions adding noise proportional to more realistic errors than the ones estimated by phot (calculated through artificial star tests). From the results of these simulations it was found by running KMM tests that it is very likely to have a bimodal distribution appearing more and more unimodal in the transition from the bluest ($(g-z)$) to the reddest ($(z-k)$) colours.   
Whether the absence of horizontal branch stars in $(z-k)$ can indeed be responsible for the weakening of bimodality in this colour as oposed to $(g-z)$ (and $(g-k)$) and support the scenario proposed by \cite[Yoon, Yi \& Lee (2006)]{yoon06} cannot be concluded with the current data set as observational uncertainties in the K-band can also be the mechanism behind this.

\begin{figure}[h]
\begin{center}
 \hspace*{1.9cm}
\includegraphics[width=1.0in]{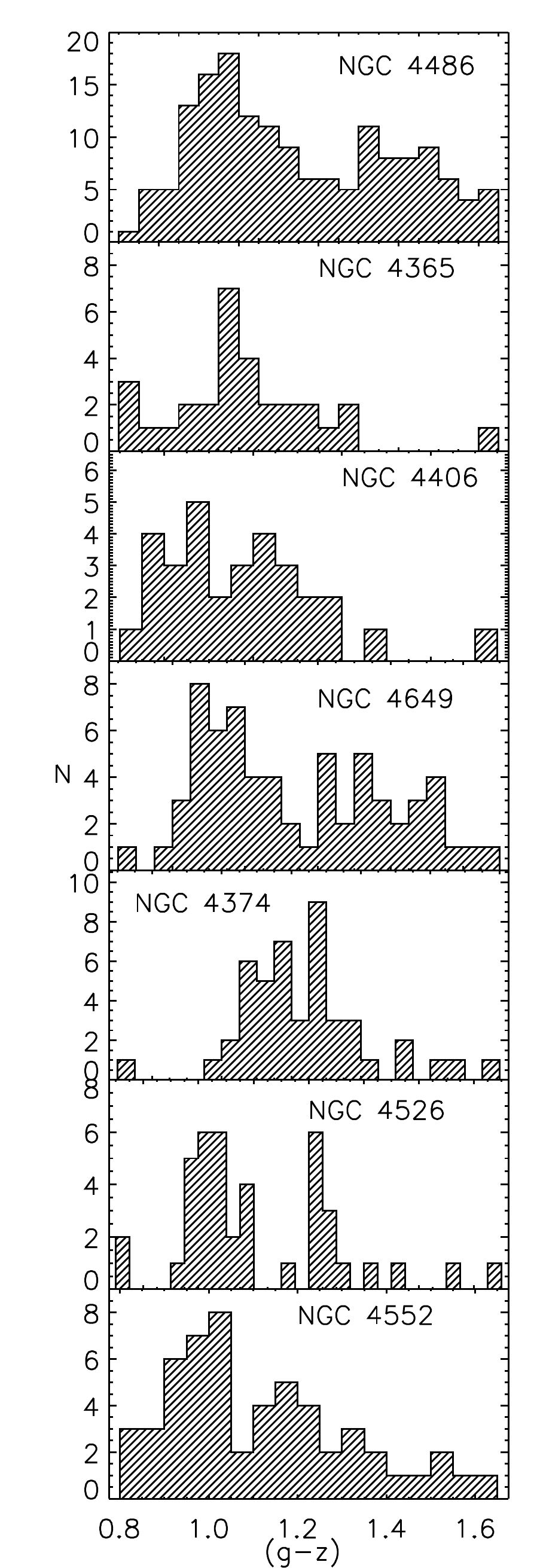} 
\includegraphics[width=1.0in]{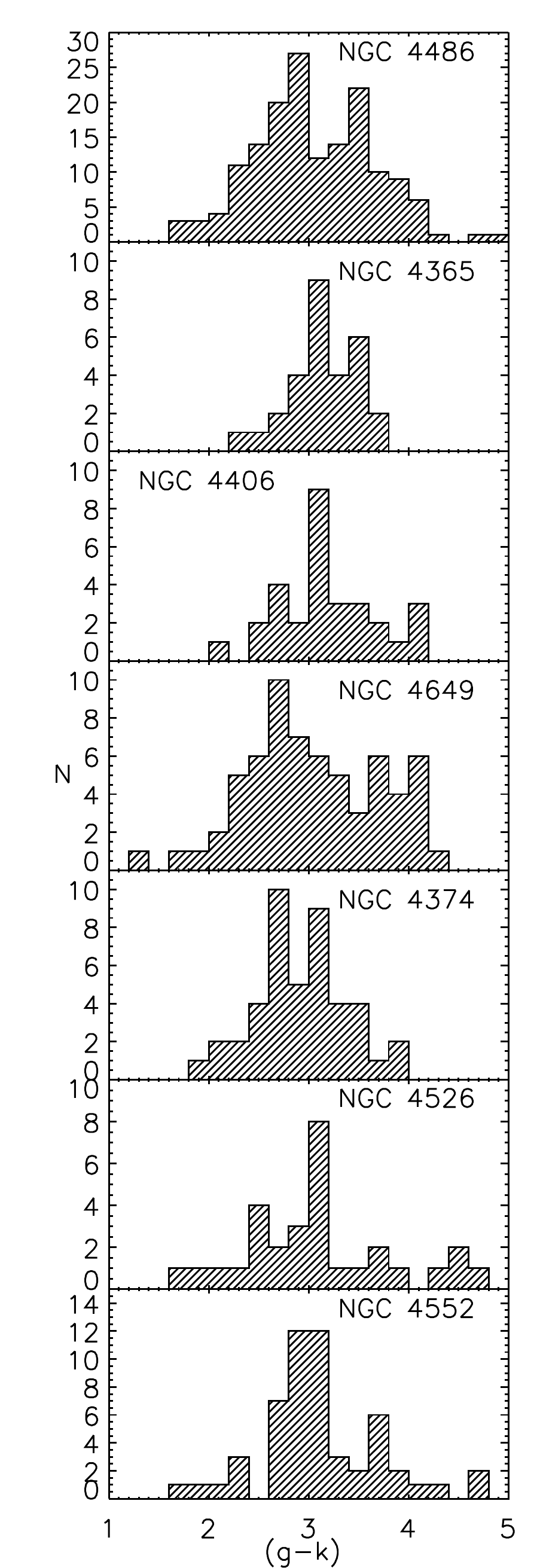} 
\includegraphics[width=1.0in]{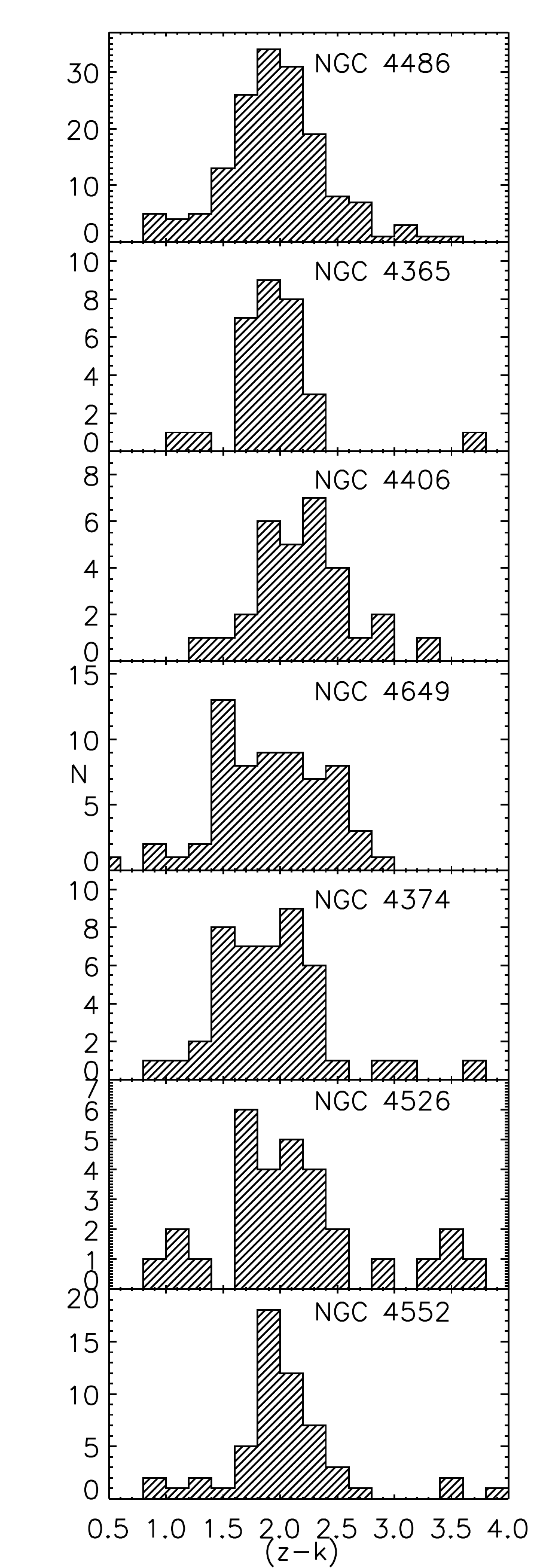} 
 \hspace*{1.9cm}
 \includegraphics[width=1.0in]{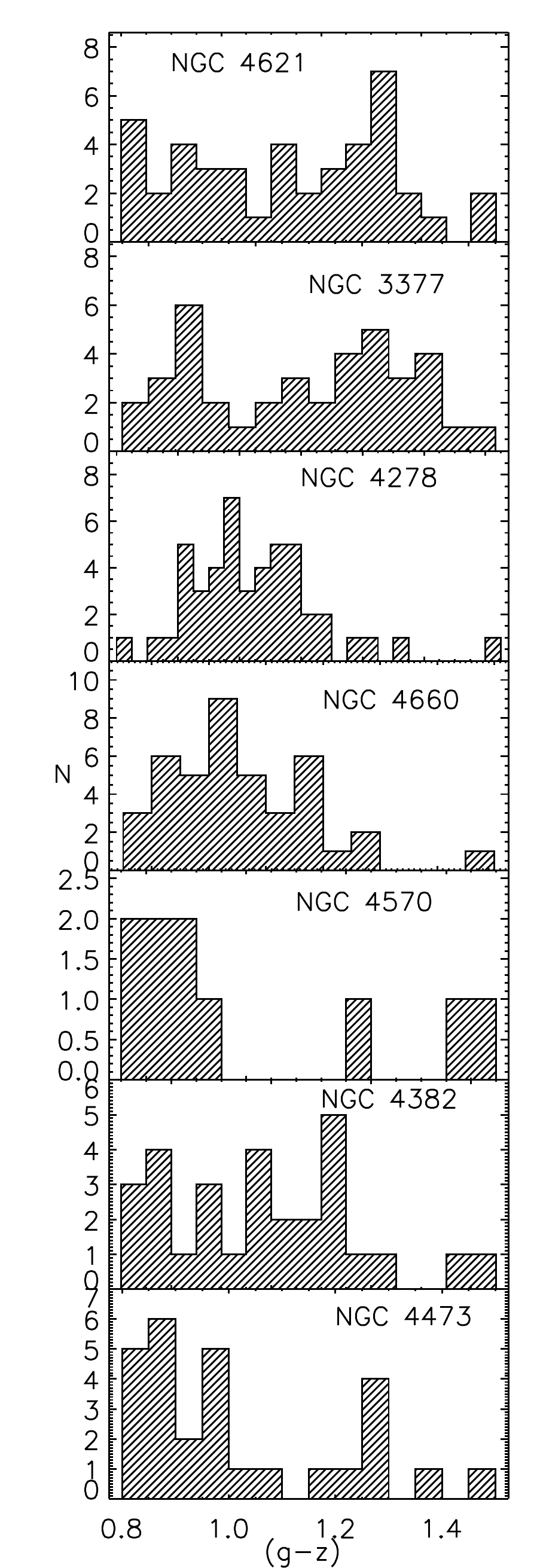}
\includegraphics[width=1.0in]{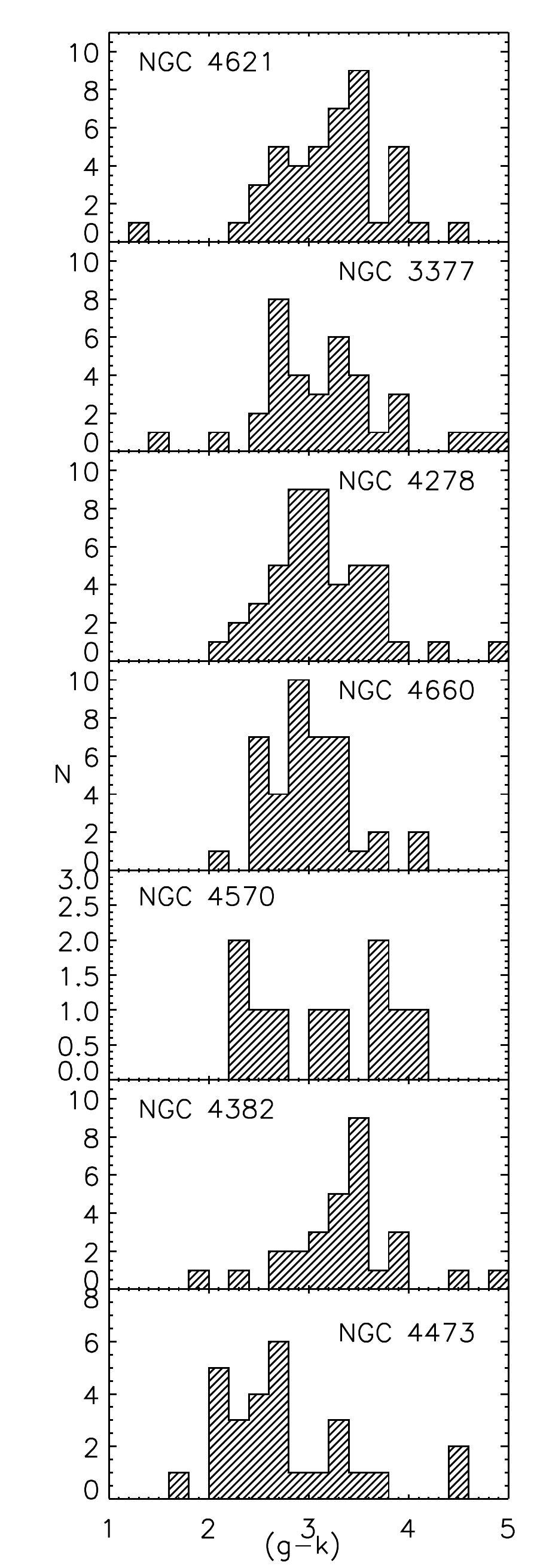}
\includegraphics[width=1.0in]{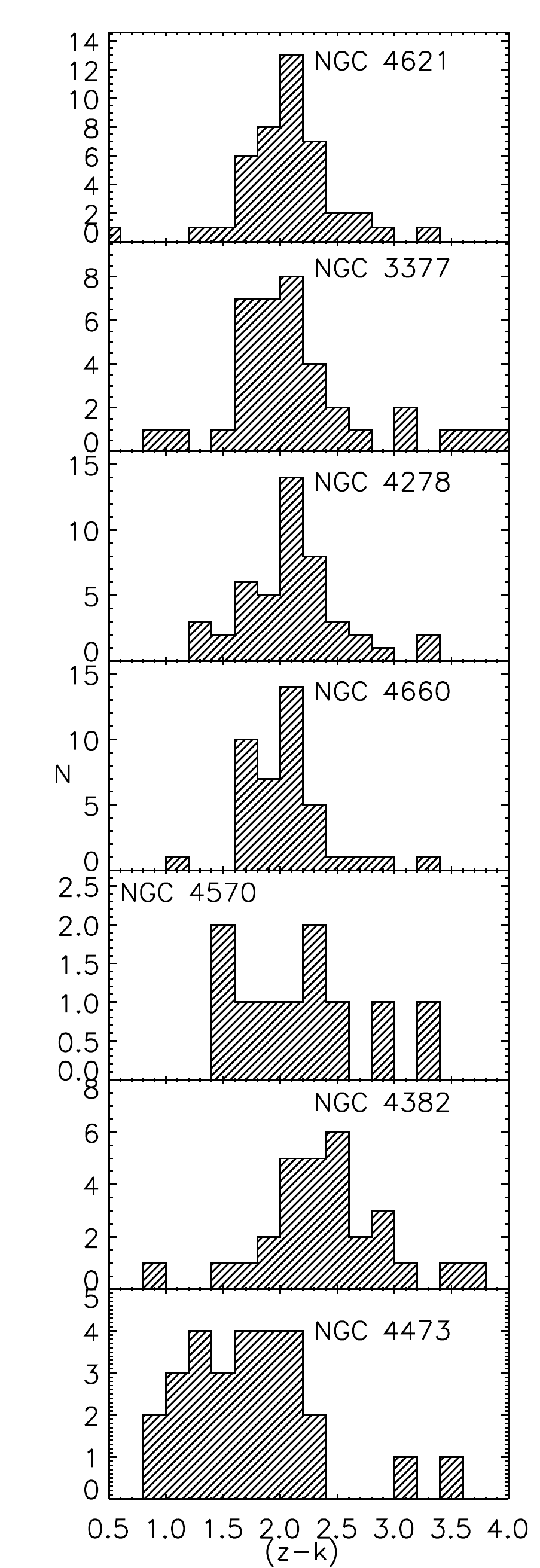}
 \caption{$(g-z)$, $(g-k)$ and $(z-k)$ colour distributions for the GC systems of the 14 E/S0s shown if Fig.\,1. Note that the bimodality becomes less evident in $(g-k)$ if compared to $(g-z)$ and even less pronounced in $(z-k)$. The latter is the colour that should have the smallest contribution from horizontal branch stars.}
   \label{Fig2}
\end{center}
\end{figure}

\subsection{Horizontal branch morphology}
In Fig.\,3 2-colour plots, $(g-k)$ \textit{vs.} $(z-k)$ for GCs of M87 and NGC 4649, indicated by different symbols are shown.  Note the wavy feature the data presents around $(g-k)\,\sim\,3.2$ and $(z-k)\,\sim\,2$. For the sake of clarity, in the left panel we show the data points alone with their corresponding error bars. A SPoT-Teramo 14 Gyr SSP with a realistic treatment of horizontal branch morphology (\cite[Raimondo et al. 2005]{raimondo05}) is over plotted in the middle panel and a Padova 14 Gyr SSP with \cite[Marigo et al. (2008)]{marigo08} isochrones is over plotted in the right one.
This wavy feature is also present in the SPot-Teramo SSP model but does not fit well with the data specially for the redder part. Other SSP models (\cite[Charlot \& Bruzual 2009]{cb09}, \cite[Maraston 2005]{m05} and Padova) that do not take into account the horizontal branch morphology do not show this behaviour as can be seen in the right panel for the case of Padova.
By comparing Fig.\,3 to Fig.\,1. of \cite[Yoon, Yi \& Lee (2006)]{yoon06} one concludes that $(z-k)$ is indeed a good indicator for [Fe/H] and that $(g-k)$ still contains some contribution of horizontal branch stars.   
Higher S/N, deeper data, sampling more GCs would be necessary to understand whether the nonlinear behaviour in the colour-metallicity relation induced by the horizontal branch morphology, as attested by Fig.\,3. would indeed drive a unimodal metallicity distribution into a bimodal (optical) colour distribution as proposed by \cite[Yoon, Yi \& Lee (2006)]{yoon06}.

\begin{figure}[h]
\begin{center}
\includegraphics[width=1\textwidth]{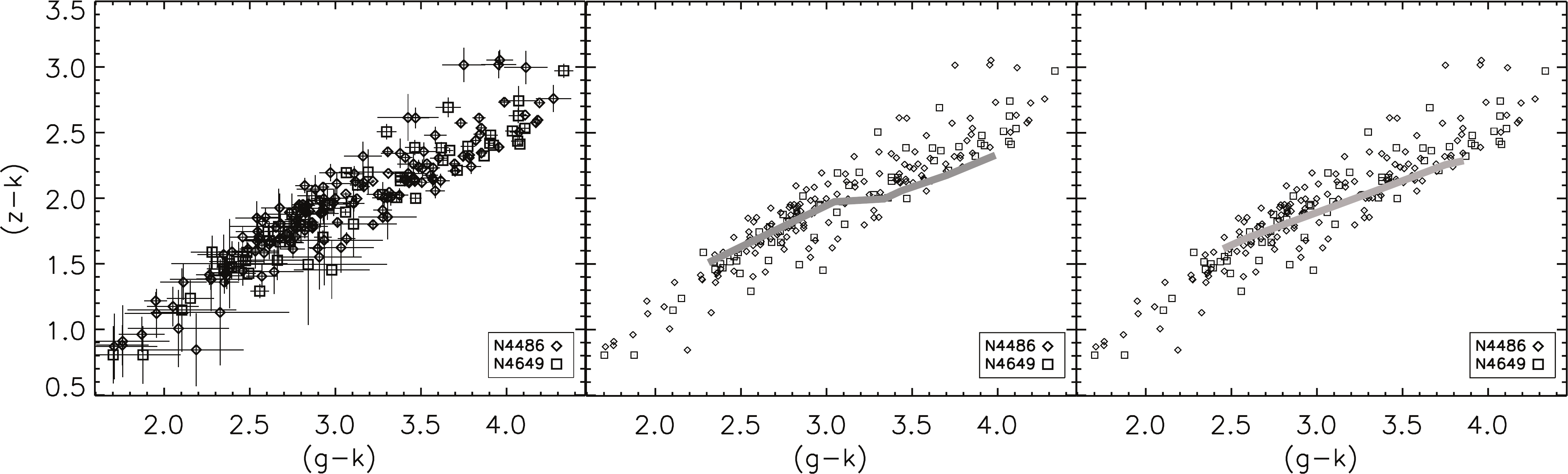}
 \caption{2-colour plots, $(g-k)$ \textit{vs.} $(z-k)$ for GCs of M87 and NGC 4649, indicated by different symbols. The error bars are only presented in the left panel for clarity. Note the wavy feature the data presents around $(g-k)\,\sim\,3.2$ and $(z-k)\,\sim\,2$. The SPot-Terramo 14 Gyr SSP with a realistic treatment of horizontal branch morphology (\cite[Raimondo et al. 2005]{raimondo05}) is overplotted in the middle panel and the Padova 14 Gyr SSP with \cite[Marigo et al. (2008)]{marigo08} isochrones is over plotted in the right panel. }
   \label{Fig3}
\end{center}
\end{figure}

\section{Conclusions}
A sample of GC systems in E/S0s was observed in the K-band with LIRIS/WHT and combined with archival ACS/HST imaging in the $g$ and $z$ bands.
From these observations we intended to study the overall ages and metallicity distributions of GC systems.

We find that age dating GCs as old or intermediate age depends on the choice of the model.
Padova SSPs for old ages ($\sim 12,14$ Gyrs) with \cite[Marigo et al. (2008)]{marigo08} isochrones with a new treatment for the TP-AGB phase fits the data well. \cite[Maraston (2005)]{m05} and \cite[Charlot \& Bruzual (2009)]{cb09} models yield intermediate ages  ($\sim 2-3$ Gyrs) for the majority of the GCs.
NGC\,4365 GC system is not found to be an exception, in general all the GC systems in the 2-colour diagrams look very much alike. NGC\,4406 is found to be the strongest candidate for having intermediate ages. 

As far as metallicity distributions are concerned the bimodality becomes less evident in $(g-k)$ if compared to $(g-z)$ and even less pronounced in $(z-k)$. The latter colour is the one that should have the smallest contribution from horizontal branch stars. The disappearance of bimodality in this colour while evident in the optical $(g-z)$ colour could be attributed to the \cite[Yoon, Yi \& Lee (2006)]{yoon06} effect although the observational uncertainties could also account for it.
The relation $(g-k)$ \textit{vs.} $(z-k)$ shows a wiggle around  $(g-k)\,\sim\,3.2$ and $(z-k)\,\sim\,2$ which is present in the 14 Gyr SPoT-Teramo SSP models where there is a proper treatment of the horizontal branch morphology. It is concluded that redder colours such as $(z-k)$ are ideal metallicity indicators.

This study calls attention to the importance of choice of SSP models. How well a set of models fits the data seems to be linked to which stellar evolutionary stages are incorporated in it and how this is done. As far as deriving overall ages of GCs with the optical/NIR imaging technique, it becomes clear that different kinds of TP-AGB treatments yield different results. SSPs with recent TP-AGB models incorporated are consistent with old GC ages. The addition or not of a realistic treatment of the HB morphology can also make a big difference.
Whether this evolutionary phase is or not responsible for optical colour bimodality cannot be answered with the data here presented.

\begin{acknowledgements}
We would like to thank G. Bruzual, S. Charlot and  M. Cantiello for kindly providing access to their models prior to publication.
\end{acknowledgements}

\end{document}